\documentclass{aa}  
\usepackage{graphicx}
\usepackage{txfonts}
\begin{document}
   \title{Near-infrared reddening of extra-galactic GMCs in a face-on geometry}


   \author{J. Kainulainen \inst{1 \and 2} \and M. Juvela\inst{2} \and J. Alves \inst{3}  
          }

   \offprints{J. Kainulainen}

   \institute{European Southern Observatory, Karl-Schwarzschild-Str. 2, D-85748 Garching bei M\"unchen\\
              \email{jouni.kainulainen@helsinki.fi}
   \and Observatory, P.O. Box 14, SF-00014 University of Helsinki
   \and  Calar Alto Observatory, Centro Astron\'omico Hispano, Alem\'an, C/q Jes\'us Durb\'an Rem\'on 2-2, 04004 Almeria, Spain
}

   \date{<>; <>}

 
  \abstract
   {}
   {We describe the near-infrared reddening signature of giant molecular clouds (GMCs) in external galaxies. In particular, we examine the $E_\mathrm{J-H}$ and $ E_\mathrm{H-K}$ color-excesses, and the effective extinction law observed in discrete GMC regions. We also study the effect of the relative scale height of the GMC distribution to the color-excesses, and to the observed mass function of GMCs when the masses are derived using color-excess as a linear estimator of mass.}
   {We perform Monte Carlo radiative transfer simulations with 3D models of stellar radiation and clumpy dust distributions, resembling a face-on geometry. The scattered light is included in the models, and near-infrared color maps are calculated from the simulated data. We perform the simulations with different scale heights of GMCs and compare the color-excesses and attenuation of light in different geometries. We extract GMCs from the simulated color maps and compare the mass functions to the input mass functions. }
   {The effective near-infrared reddening law, i.e. the ratio $E_\mathrm{J-H}/E_\mathrm{H-K}$, has a value close to unity in GMC regions. The ratio depends significantly on the relative scale height of GMCs, $\xi$, and for $\xi$ values $0.1 \dots 0.75$ we find the typical ratios of $0.6 \dots 1.1$. The effective extinction law turns out to be very flat in GMC regions. We find the ratios of apparent ectinctions of $A_\mathrm{H}^\mathrm{a}/A_\mathrm{K}^\mathrm{a}=1.35\dots1.55$ and $A_\mathrm{J}^\mathrm{a}/A_\mathrm{H}^\mathrm{a}=1.15$. The effect of the scattered flux on the effective reddening law, as well as on the effective extinction law, is significant. Regarding the GMC mass function, we find no correlation between the input and observed slopes of the mass functions. Rather, the observed slope reflects the parameter $\xi$ and the dynamical range of the mass function.
As the observed slope depends on the geometric parameters which are not known, it is not possible to constrain the slope of the mass function using this technique. We estimate that only a fraction of $10\dots20\%$ of the total mass of GMCs is recovered, if the observed color-excess values are transformed to masses using the Galactic reddening law. In the case of individual clouds the fraction can vary between $\sim 0\dots50\%$.}
   {}

   \maketitle
%

\section{Introduction}


The dust content of galaxies has an enormous effect on their observed properties in both local and high-redshift universe. The dust is responsible for the thermal emission at infrared wavelengths, it attenuates the stellar radiation from near-infrared to optical, and blocks efficiently radiation at ultraviolet wavelengths. The realistic description of the effects of dust is thus crucial for numerous applications ranging from the description of galaxy evolution to the estimation of the amount of dark matter. Albeit important, the estimation of the impact of dust is, in most of the cases, all but a trivial task. This is because the effects caused by dust depend drastically, not only on the amount of it, but also on the exact spatial distribution and the physical composition of it.


Perhaps the most obvious manifestation of dust in galaxies outside the Milky Way are the lanes of dark clouds seen in the nearby spiral galaxies. In face-on spirals the distribution of these dusty giant molecular clouds (GMCs) follows qualitatively the stellar spiral arms. In edge-on spirals the GMCs are seen as an optically thick layer, concentrated heavily on the midplane of the galaxy. At the distance of nearby galaxies the sizes of GMCs correspond to angular diameters of a few arcseconds. This basically rules out the possibility to study them in detail using single-dish observations at frequencies of common tracers of molecular gas, such as CO. Interferometric observations can achieve the required resolution, but mapping large areas is overwhelmingly time consuming and sensitivity is rather poor.


Direct imaging of galaxies from optical to near-infared offers a high-resolution view to the distribution of GMCs inside them. At these wavelengths an arcsecond-scale resolution is easily achieved with standard imaging observations. Qualitatively, the presence and distribution of GMCs can be established by color maps made from observations in two broadband filters, e.g. $B-V$ or $J-K$. This approach has been used, often accompanied with image processing techniques, such as unsharp masking, to disentangle the GMCs from the color maps (e.g. Regan \cite{regan95}, \cite{regan00}; Howk \& Savage \cite{howk97}, \cite{howk99}, \cite{howk00}; Trewhella \cite{trewhella98}; Elmegreen \cite{elmegreen98}; Thompson et al. \cite{thompson04}). The studies are typically performed in optical wavelenghts due to the higher contrast towards the dust features, and until the last decade, due to the lack of large near-infrared array cameras.


The interpretation of the observed color map data is severely hampered by the unknown relative geometry of dust and stars in the galaxy. The flux emitted by stars between a dust feature and the observer, as well as the star light that enters a cloud from all directions and scatters towards the observer, both reduce the reddening signature of dust clouds. Thus, the description of reddening effects requires radiative transfer calculations. As the basic problem regarding the radiative transfer in dusty medium is important for several fields, the studies of dusty systems have been made in the past by numerous authors and from several points of view. In the context of embedded, galaxy-like dust distributions, a lot of emphasis has been placed on describing the obscuration by inhomogenous dust which has a low volume filling factor, and in which large density contrasts are present (e.g. Witt et al. \cite{witt92}, \cite{witt96}, \cite{witt00}; Pierini et al. \cite{pierini04}). This set-up resembles the geometry of mixed GMCs and stellar distribution, with low column density lines of sight dominating the light escaping from the system (i.e. the observed flux). It is particularly tailored to interpret observations of central regions of starburst galaxies, or high redshift star-forming galaxies. It is a common result of these studies that color-excesses can be highly nonlinear functions of optical depth and they saturate at relatively low values. The small scale structure, i.e. the clumpiness, significantly lowers the observed reddening compared to that caused by a homogenous dust distribution, and the effect of scattering is not neglible at most of the wavelenghts.

In this paper we examine the obscuration effects observed in the regions of particularly high column density, i.e. in the GMCs. The configuration chosen in our simulations differs from the studies quoted above, as in our models the dust structures experience strong illumination from an extended distribution of stellar light surrounding the dust. The dust clouds are small compared to the extent of the stellar light distribution, and therefore the flux observed in on-cloud pixels is likely to be dominated by the emission of the stars completely in front of all the dust. The paper is second in series where we study the near-infrared reddening properties of GMCs, which are embedded inside a disk-like stellar radiation field. In the first paper we showed that the slope of the observed GMC mass function, derived from the NIR color-excess data using foreground screen approximation, can significantly differ from the true mass function (Kainulainen et al. \cite{kainulainen07}, Paper I hereafter). In this paper we expand the results of Paper I and describe the near-infrared reddening and attenuation occuring in the GMC regions in detail. We examine the effective reddening and extinction laws in our models, with emphasis on how the relative scale height of the GMC distribution, $\xi$, affects their values. We also examine the shape of the observed mass function as a function of the parameter $\xi$ to find out, if the slope of the mass function can be constrained when some geometric parameters are known. In \S \ref{sec_modeling} the radiative transfer method and cloud models are presented. The results are presented and discussed in \S\ref{sec_results} and \S\ref{sec_discussion}. In \S\ref{sec_conclusions} we give our conclusions.

\section{Modeling}    

\label{sec_modeling}

\begin{figure}

  \centering

  \includegraphics[width=0.99\columnwidth]{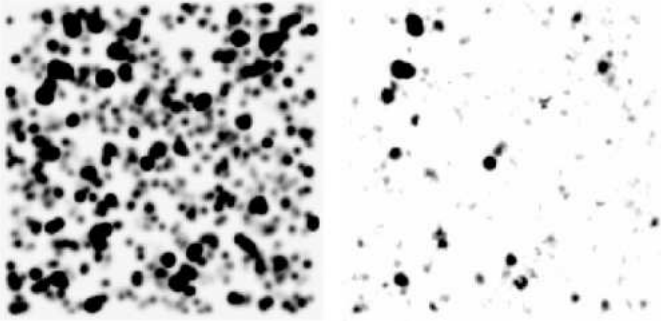}

  \caption{An example of one realization of a model used in the simulations (model \#14). {\bf Left: }Column density. {\bf Right: }Corresponding simulated $H-K$ color map.}
  \label{fig_models}

\end{figure}

The radiative transfer method, the basic geometry of the models, and the cloud extraction has been described in Paper I. In short, the models consist of an exponential distribution of discrete clouds, which are embedded in an exponential distribution of stellar light. The left panel of Fig. \ref{fig_models} shows an example of input density distribution projected to a plane. The distributions of clouds and stars are characterized by their scale heights, $z_\mathrm{c}$ and $z_\mathrm{s}$, respectively. The ratio of the scale heights, $\xi= z_\mathrm{c}/z_\mathrm{s}$, is the main parameter describing the model geometry. The models are viewed from a face-on angle, i.e. the inclination angle of zero. The dust properties in the simulations are based those of Draine (\cite{draine03}), and we use the tabulated values for Milky Way and R$_\mathrm{V}=3.1$, which are available on the web\footnote{http://www.astro.princeton.edu/$\sim$draine/dust/}.

We generated 36 models, which result from choosing four different values for $\xi$, three values for the slope of the input mass function, $\alpha_\mathrm{init}$, and three values for the low-mass cut-off of the mass function, $M_\mathrm{low}$. The models and corresponding input parameters are listed in table \ref{tab_results}.  In addition to the three parameters above, table \ref{tab_results} gives the volume filling factor, and the surface density filling factor in each model. The filling factor is defined as the number of cells with non-zero density divided by the total number of cells. The surface density filling factor is defined in a similar way, but it is calculated after projecting the model on a two-dimensional plane.

The radiative transfer calculations are performed in $JHK$ bands, and the resulting fluxes are used to construct simulated surface brightness maps. The surface brightness maps are further combined to color maps, namely $H-K$ and $J-H$. The right panel of Fig. \ref{fig_models} shows an example of resulting $H-K$ color map of one realization of model \#14. The color maps are further combined to $A_\mathrm{V}$ extinction maps using the NICER method (Lombardi \& Alves \cite{lombardi01}), applied to the surface brightess data. In the NICER method, the estimate of $A_\mathrm{V}$ results from the comparison of the colors of on-cloud pixels to the colors of nearby off-cloud pixels. In our models the colors of off-cloud regions are constant, and thus we selected one off-cloud region per model from which the off-cloud colors were calculated. In the selection we used the empty regions of $J-H$ color maps. The individual clouds are identified from the extinction maps using the clumpfind routine 'clfind2d' (Williams et al. \cite{williams94}). The masses of identified clouds are calculated by summing up the extinction of all the on-cloud pixels, assuming the foreground screen geometry, and using the Milky Way ratio of $N(H2)/A_\mathrm{V}=9.4 \times 10^{20}$ cm$^{-2}$mag$^{-1}$ (Bohlin et al. \cite{bohlin78}). Thus, the masses we refer to in the paper represent the gaseous masses of the clouds, and we express it in the units of solar masses, M$_\odot$. The masses are then used to construct the observed mass functions presented in Section \ref{sec_results}. It is noteworthy that practically all the clouds that were embedded in the models ($\gtrsim 95$ \%) were dense enough to be detected, if they were located completely in front of the stellar light distribution. This means that there is no incompleteness in the observed mass function due to the family of clouds that intrinsically \emph{cannot} be detected.

The same procedure of cloud extraction was used to construct the true cloud mass functions from the inital density distributions that were projected on a two-dimensional plane. The power-law model was fitted to the data to extract the slope of the true mass function, $\alpha_{true}$. We note that this is not exactly the same as $\alpha_\mathrm{init}$, which was defined to be the slope of the distribution from which the masses of input clouds were determined. In other words, the true mass functions presented in Section \ref{sec_results} are not composed of the initial input masses of the clouds, but rather their masses after 2D projection and the clumpfind treatment. This was done to make sure that the cloud extraction procedure itself does not affect the slope of the observed mass function. The results of the power-law fits show that the slopes of the true mass functions are in good agreement with the initial ones, deviating 0.2 from the value of them at most (the slopes are given in columns 2 and 7 of Table \ref{tab_results}).

\section{Results}        
\label{sec_results}

The results of the radiative transfer simulations, which are the main results of this paper, are summarized in Table \ref{tab_results}. The first five columns of Table \ref{tab_results} give the input parameters for each model. The definitions of these parameters have been given in Section \ref{sec_modeling}. The following columns show the output properties of the models. $\alpha_\mathrm{true}$ and $\alpha_\mathrm{obs}$, given in columns 6 and 7, are the exponents of the power-law fits to the true and observed mass functions. The 8th column gives the fraction of the detected clouds, i.e. the number of clouds extracted from the NICER extinction map divided by the number of input clouds. The column 9 shows the total observed mass divided by the the total input mass. The columns 10 and 11 show the slope of the fit made to the frequency distribution of color-excess pixel values (see Section \ref{sec_color-excesses}).

\begin{table*}
\begin{minipage}[t]{\columnwidth}
\caption{The input parameters and observed properties of the models.}
\label{tab_results}
\centering
\renewcommand{\footnoterule}{}  
\begin{tabular}{c c c c c c c c c c c c}     
\hline\hline       
                      
    & \multicolumn{5}{c}{Input parameters} & & \multicolumn{4}{c}{Observed properties} \\
\hline
$\#$  & $\alpha_\mathrm{init}$ & $\xi$ & $\log M_\mathrm{low}$ [M$_\odot$] & $ff$ [$\%$]& $\rho_\mathrm{2D}$ [$\%$]& $\alpha_\mathrm{true}$ & $\alpha_\mathrm{obs}$ & $\frac{N_\mathrm{obs}}{N_\mathrm{tot}}$ [$\%$]& $\frac{M_\mathrm{obs}}{M_\mathrm{tot}}$ [$\%$] & $A_\mathrm{H-K}$\footnote{Slope of the fit to the frequency distribution of color-excess pixel values, i.e. $\log y=-Ax+B$.} & $A_\mathrm{J-H}$ \\
\hline

   1\footnote{With the parameter settings of this run only a very small fraction of the clouds were detected.}  & 2.5 & 0.10 & 4.0 & 23  & 11       & -  & -  & $<1$ & -  
& -  $\pm$ -   & -  $\pm$ -   \\  
   2  & 2.5 & 0.27 & 4.0 & 12  & 11       & 2.6 & 2.6 & 2.7 & 7.3 
& 6.7 $\pm$ 0.75 & 9.0 $\pm$ 0.33 \\  
   3  & 2.5 & 0.50 & 4.0 & 7.2 & 11       & 2.6 & 2.2 & 6.2 & 11  
& 6.3 $\pm$ 0.78 & 5.8 $\pm$ 0.30 \\  
   4  & 2.5 & 0.75 & 4.0 & 4.9 & 11       & 2.4 & 2.2 & 8.6 & 22  
& 3.7 $\pm$ 0.49 & 5.2 $\pm$ 0.30 \\  

   5  & 2.0 & 0.10 & 4.0 & 13  & 14       & 1.9 & 2.7 & 7.1 & 5.8 
 & 10  $\pm$ 0.40 & 9.0 $\pm$ 0.26 \\  
   6  & 2.0 & 0.27 & 4.0 & 12  & 13       & 2.0 & 2.7 & 7.9 & 7.2 
 & 7.1 $\pm$ 0.26 & 7.4 $\pm$ 0.24 \\  
   7  & 2.0 & 0.50 & 4.0 & 7.2 & 14       & 1.9 & 2.0 & 11  & 8.8 
& 5.7 $\pm$ 0.21 & 5.2 $\pm$ 0.21 \\  
   8  & 2.0 & 0.75 & 4.0 & 5.4 & 14       & 2.0 & 2.0 & 14  & 9.8 
& 3.7 $\pm$ 0.13 & 4.5 $\pm$ 0.13 \\  

   9  & 1.5 & 0.10 & 4.0 & 28  & 17       & 1.5 & 2.8 & 33  & 7.0 
& 8.4 $\pm$ 0.23 & 8.3 $\pm$ 0.26 \\  
   10 & 1.5 & 0.27 & 4.0 & 14  & 18       & 1.5 & 2.4 & 34  & 6.3 
& 7.6 $\pm$ 0.25 & 7.4 $\pm$ 0.11 \\  
   11 & 1.5 & 0.50 & 4.0 & 9.0 & 18       & 1.6 & 2.0 & 35  & 8.3 
& 5.1 $\pm$ 0.13 & 5.4 $\pm$ 0.13 \\  
   12 & 1.5 & 0.75 & 4.0 & 6.2 & 17       & 1.5 & 2.1 & 33  & 11  
& 4.1 $\pm$ 0.11 & 4.5 $\pm$ 0.08 \\  %

\hline

   13 & 2.5 & 0.10 & 4.5 & 28  & 24       & 2.6 & 3.1 & 8.0 & 7.0 
& 9.5 $\pm$ 0.85 & 10  $\pm$ 0.41 \\
   14 & 2.5 & 0.27 & 4.5 & 18  & 24       & 2.6 & 2.7 & 14  & 8.4 
& 7.0 $\pm$ 0.68 & 7.7 $\pm$ 0.65 \\
   15 & 2.5 & 0.50 & 4.5 & 12  & 24       & 2.7 & 2.8 & 24  & 10  
& 5.1 $\pm$ 0.43 & 6.2 $\pm$ 0.34 \\
   26 & 2.5 & 0.75 & 4.5 & 8   & 24       & 2.7 & 2.4 & 31  & 14  
& 3.6 $\pm$ 0.34 & 4.6 $\pm$ 0.33 \\


   17 & 2.0 & 0.10 & 4.5 & 55  & 31       & 2.1 & 2.4 & 19  & 8.2 
& 9.2 $\pm$ 0.35 & 9.5 $\pm$ 0.30 \\ 
   18 & 2.0 & 0.27 & 4.5 & 38  & 31       & 2.0 & 2.2 & 21  & 9.2 
& 6.8 $\pm$ 0.23 & 7.2 $\pm$ 0.20 \\ 
   19 & 2.0 & 0.50 & 4.5 & 27  & 31       & 2.0 & 2.0 & 27  & 13  
& 4.9 $\pm$ 0.20 & 5.2 $\pm$ 0.20 \\ 
   20 & 2.0 & 0.75 & 4.5 & 20  & 31       & 1.9 & 2.0 & 34  & 13  
& 4.1 $\pm$ 0.13 & 4.0 $\pm$ 0.12 \\ 

   21 & 1.5 & 0.10 & 4.5 & 27  & 17       & 1.5 & 2.4 & 17  & 7.2 
& 9.4 $\pm$ 0.27 & 8.2 $\pm$ 0.19 \\  
   22 & 1.5 & 0.27 & 4.5 & 14  & 15       & 1.5 & 2.3 & 9.8 & 8.5 
& 6.9 $\pm$ 0.25 & 6.8 $\pm$ 0.20 \\  
   23 & 1.5 & 0.50 & 4.5 & 10  & 16       & 1.5 & 2.4 & 13  & 8.0 
& 4.6 $\pm$ 0.15 & 4.4 $\pm$ 0.20 \\  
   24 & 1.5 & 0.75 & 4.5 & 8.0 & 19       & 1.5 & 2.5 & 19  & 9.1 
& 3.9 $\pm$ 0.08 & 3.8 $\pm$ 0.09 \\  

\hline


   25 & 2.5 & 0.10 & 5.0 & 36  & 32       & 2.7 & 2.7 & 53  & 5.1 
 & 12  $\pm$ 0.57 & 11  $\pm$ 0.19 \\  
   26 & 2.5 & 0.27 & 5.0 & 20  & 32       & 2.5 & 2.6 & 56  & 6.7 
& 8.0 $\pm$ 0.48 & 9.3 $\pm$ 0.22 \\  
   27 & 2.5 & 0.50 & 5.0 & 10  & 32       & 2.5 & 2.8 & 58  & 8.1 
 & 7.1 $\pm$ 0.23 & 6.4 $\pm$ 0.17 \\  
   28 & 2.5 & 0.75 & 5.0 & 7.8 & 32       & 2.5 & 2.6 & 59  & 11  
& 4.3 $\pm$ 0.17 & 4.5 $\pm$ 0.14 \\  

   29 & 2.0 & 0.10 & 5.0 & 37  & 30       & 2.0 & 2.9 & 68  & 7.2 
& 11  $\pm$ 0.35 & 11  $\pm$ 0.28 \\ 
   30 & 2.0 & 0.27 & 5.0 & 14  & 24       & 2.0 & 2.7 & 91  & 9.0 
& 7.6 $\pm$ 0.37 & 7.5 $\pm$ 0.31 \\ 
   31 & 2.0 & 0.50 & 5.0 & 11  & 30       & 2.0 & 2.3 & 83  & 10  
& 5.6 $\pm$ 0.15 & 5.5 $\pm$ 0.23 \\ 
   32 & 2.0 & 0.75 & 5.0 & 7.0 & 30       & 1.9 & 2.3 & 94  & 14  
& 4.2 $\pm$ 0.11 & 4.3 $\pm$ 0.10 \\ 

   33 & 1.5 & 0.10 & 5.0 & 28  & 17       & 1.4 & 2.4 & 71  & 7.0 
& 8.2 $\pm$ 0.27 & 8.1 $\pm$ 0.29 \\  
   34 & 1.5 & 0.27 & 5.0 & 14  & 18       & 1.5 & 2.2 & 78  & 7.6 
& 7.8 $\pm$ 0.19 & 7.6 $\pm$ 0.18 \\  
   35 & 1.5 & 0.50 & 5.0 & 9.0 & 18       & 1.6 & 1.9 & 85  & 8.5 
& 5.4 $\pm$ 0.12 & 5.6 $\pm$ 0.16 \\  
   36 & 1.5 & 0.75 & 5.0 & 6.2 & 17       & 1.5 & 2.1 & 78  & 10  
& 4.3 $\pm$ 0.12 & 4.6 $\pm$ 0.09 \\  

\hline                  
\end{tabular}
\end{minipage}
\end{table*}

\subsection{The observables}                              
\label{subsec_properties}


In the following we describe the observable characteristics of a typical series of simulations, and the dependency of the observables on the model parameters. We use the models 5-8 as examples in most of the figures, as they represent an average set of input parameters.

\subsubsection{Color-excesses and the effective reddening law}
\label{sec_color-excesses}

\begin{figure}
  \centering

  \includegraphics[width=0.99\columnwidth]{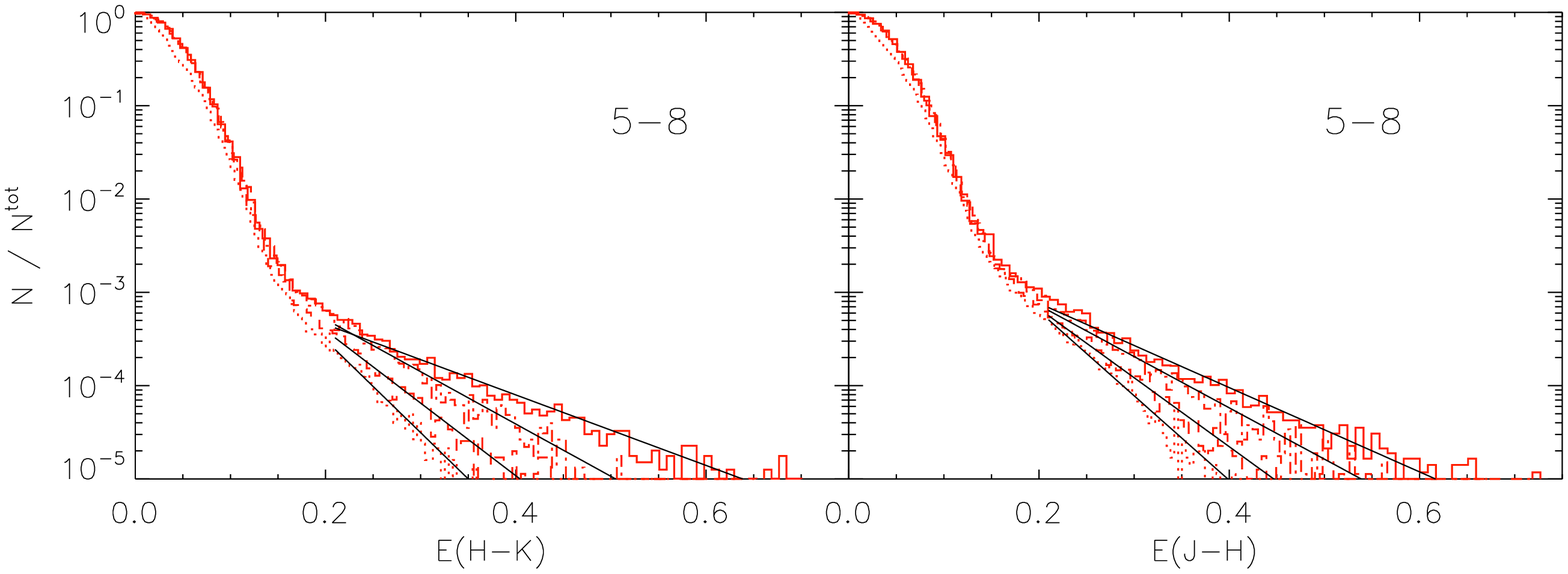}
  \includegraphics[width=0.95\columnwidth]{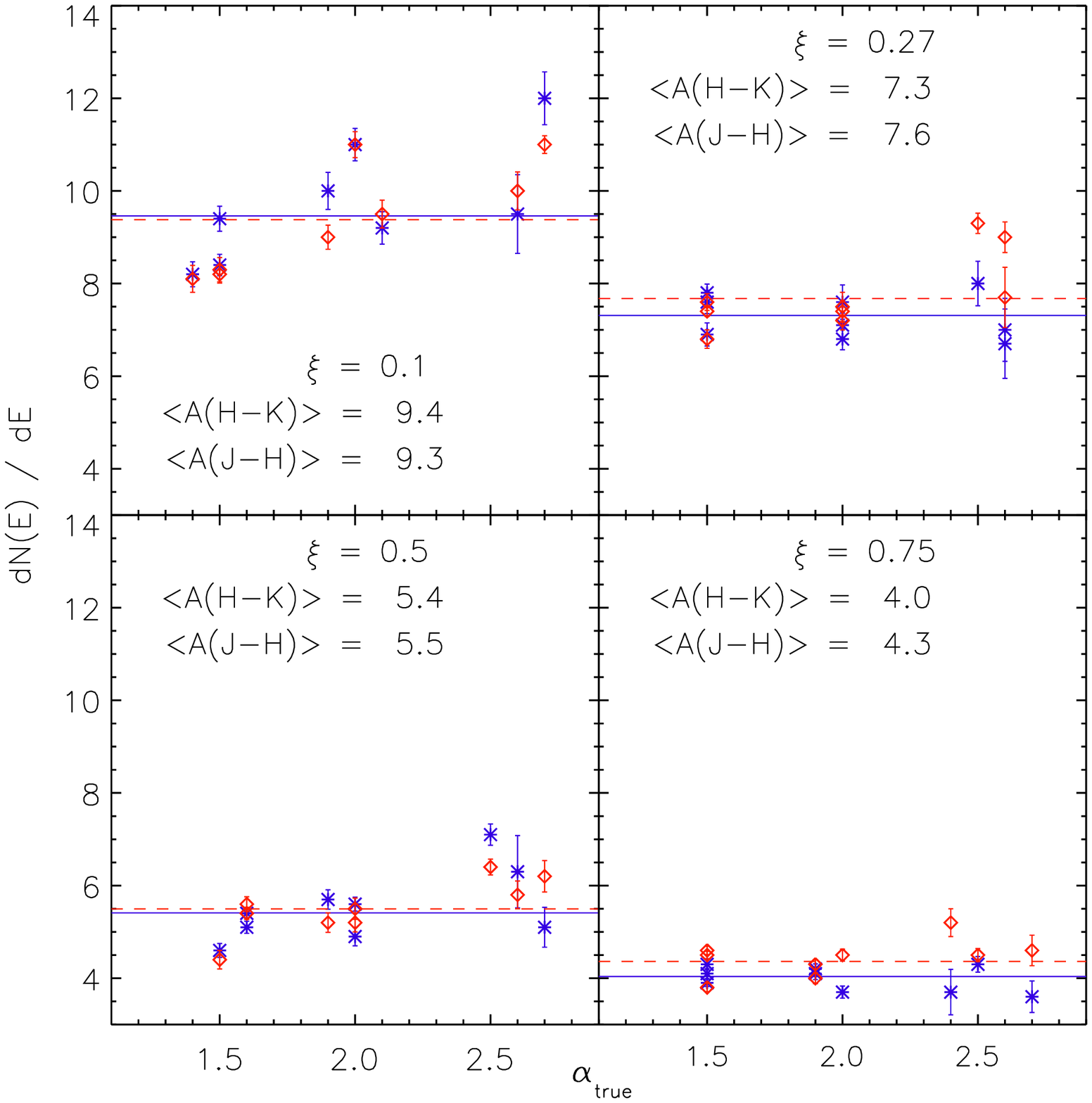} 

  \caption{{\bf Top: } Normalized histograms of $E_\mathrm{H-K}$ and $E_\mathrm{J-H}$ values in the models 5-8 (dotted, dashed, dash-dotted, and solid lines, respectively). The solid lines are fits of the equation $\log y = -Ax + B$ to the data with $E>0.2$. {\bf Bottom: }Summary of the fitted $A$ parameters for all the models of the study. The blue asterisks correspond to $E_\mathrm{H-K}$ values and the red diamonds to $E_\mathrm{J-H}$ values. The error bars represent the 1$\sigma$ error of the fits. The horizontal lines give the error-weighted average of the data points. The average values are also printed into the frames. }
         \label{fig_E-histograms}
\end{figure}

Fig. \ref{fig_E-histograms} shows the normalized frequency distributions of $E_\mathrm{H-K}$ and $E_\mathrm{J-H}$ values in models 5-8. The dotted, dashed, dash-dotted, and solid lines are for models with $\xi$ values of 0.1, 0.27, 0.5 and 0.75, respectively. Below $\sim 0.1$ the distributions are dominated by the pixel noise. At higher color-excesses the histograms follow roughly an exponential distribution, with the slope obviously depending on the value of $\xi$. The exponential decay of the color-excess values is interesting, as the initial distribution of cell densities, and thus the optical dephts of pixels, was a power-law. We fit a model $\log y = -Ax+B$ to the color-excess values higher than 0.2. We note that the change in the number of clouds per unit volume, i.e. the change in surface density, only shifts the distribution in vertical direction, but does not affect the parameter $A$ of the fit. The top panel of Fig. \ref{fig_E-histograms} shows also the fits on top of the pixel histograms. The $A$ parameters resulting from the fit are given in Table \ref{tab_results}, and summarized in the lower panel of Fig. \ref{fig_E-histograms}. The $A$ parameters depend strongly on $\xi$, being about $10$ for models with $\xi=0.1$, and shallowing with increasing values of $\xi$. In the models with $\xi=0.75$ the average value is $\sim 4$.

The ratio of the two color-excess, $E_\mathrm{J-H} / E_\mathrm{H-K}$, is especially interesting, as it describes \emph{the effective reddening law} in the particular geometry. We fitted a linear model to all $(E_\mathrm{H-K},E_\mathrm{J-H})$ pixel values of a model, and the resulting slope was $\sim$1 in practically all models. This is, as expected for embedded dust distribution, much flatter than the reddening laws for Galactic dust and foreground screen geometry (in our models: $E_\mathrm{J-H} / E_\mathrm{H-K}=1.5$, Draine \cite{draine03}).

\begin{figure}
   \centering
   \includegraphics[width=0.99\columnwidth]{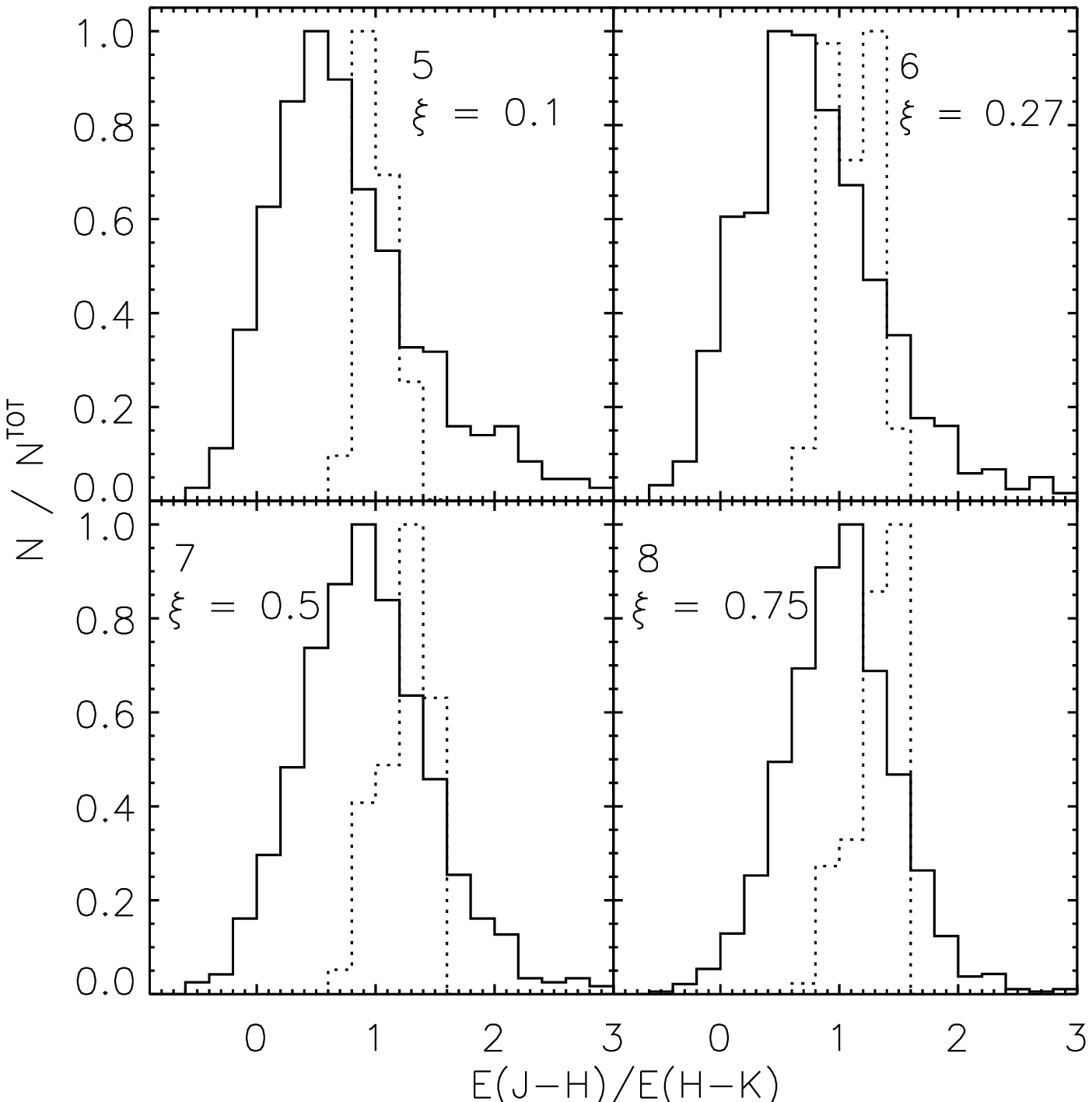} 
    \caption{The histograms of the $E_\mathrm{J-H} / E_\mathrm{H-K}$ ratios of all detected GMCs in the models 5-8. The ratio is measured at the point of maximum $E_\mathrm{H-K}$ in each cloud. The approximate peak values corresponding the $\xi$ values of 0.1, 0.27, 0.5, and 0.75 are 0.5, 0.6, 0.9, and 1.1. The number of the model and the corresponding $\xi$ value are marked into the frames. The dotted lines show the ratio if the scattered flux is neglected (see section \ref{sec_results-first}).}
         \label{fig_reddening-law}
   \end{figure}

The ratio of color-excesses can be determined individually in each of the clouds. We perform this by taking the average value of $E_\mathrm{H-K} / E_\mathrm{J-H}$ ratios of ten pixels with highest $E_\mathrm{H-K}$ to represent the ratio in that cloud. Fig. \ref{fig_reddening-law} shows the histograms of the resulting ratios in models 5-8. Even though the distributions are wide, reflecting the large scatter in the $E_\mathrm{J-H} / E_\mathrm{H-K}$ values of pixels due to photometric errors, they are clearly distinct. The peak of the distribution moves towards higher values for higher values of $\xi$, which is expected as the fraction of foreground light gets smaller on average. This behaviour remains hidden in the average relation, where the distribution is dominated by the pixels with relatively low optical depths. However, it becomes detectable when determined separately for individual clouds.  

\subsubsection{The cloud mass functions}



The masses of detected clouds are used to construct the observed mass functions, and a power-law model, described by the parameter $\alpha_\mathrm{obs}$, is fitted to data. For the binning we use the equation $N_\mathrm{bins} = (2\times n)^{1/3}$ where $n$ is the number of clouds in the histogram.  The lowest mass bin which is taken into the fit is selected individually for each model by eye. The slopes of both true and observed mass functions are listed in columns 6 and 7 of Table \ref{tab_results}. 
The appearance of the mass functions is similar to those described already in Paper I (Fig. 2). The dynamical range of the observed mass function is approximately two decades, starting from $\sim 10^{3.5}$ M$_\odot$ and terminating at $10^{5.5}$ M$_\odot$. In some models the power-law does not represent the observed mass function well over the whole observed dynamical scale. Rather, the mass function has a break-point at $M\sim10^{4.5}$ M$_\odot$, below which the mass function flattens, or is completely flat. This typically happens in the models where both $\xi$ and the low-mass cut-off of the mass function, $M_\mathrm{low}$, are high.

\begin{figure}
   \centering

   \includegraphics[width=0.95\columnwidth]{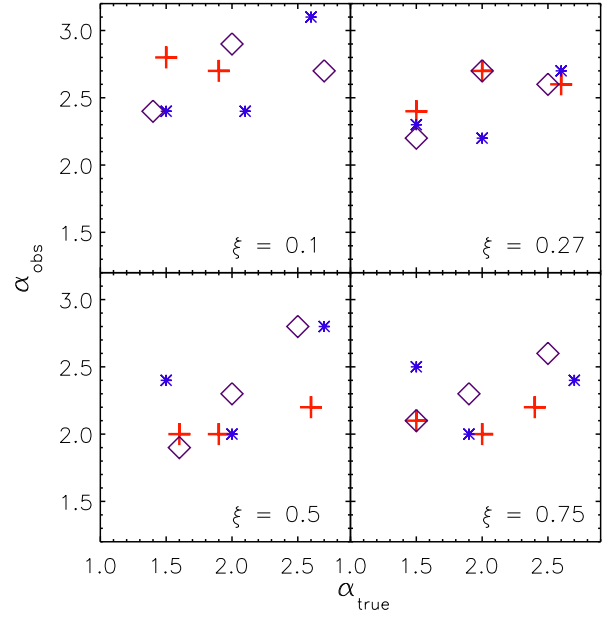} 
   \caption{The slope of the observed mass function as a function of the slope of the true mass function in all models of the study. The different panels correspond to different values of $\xi$ parameter. The red plus signs correspond to $M_\mathrm{low}=4.0$ (models 1-12), blue asterisks to $M_\mathrm{low}=4.5$ (models 13-24), and purple diamonds to $M_\mathrm{low}=5.0$ (models 25-36).}
   \label{fig_a_in_vs_a_out}

   \end{figure}

Fig. \ref{fig_a_in_vs_a_out} shows the relation between the true and observed slopes in all the models. The observed slopes vary between 1.8 and 2.9, and there is no correlation between the true and observed slopes. The models with flat input slopes are observed to have similar range of slopes as the models with steep input slopes. In some particular parameter settings, such as models marked with purple diamonds (models 27-28, 31-32, and 35-36), the observed slopes follow loosely the true slopes if the $\xi$ is high. In these models the $M_\mathrm{low}$ parameter is $10^5$ M$_\odot$. The observed slopes correlate slightly with the parameter $\xi$, as on average, the models with low $\xi$ values tend to show steeper slopes ($\alpha_\mathrm{obs} \sim 2.5$) than the models with high $\xi$ values ($\alpha_\mathrm{obs} \sim 2$). As the correlation between $\alpha_\mathrm{true}$ and $\alpha_\mathrm{obs}$ is very poor, it is not possible to estimate $\alpha_\mathrm{true}$ by using simply a measurement of $\alpha_\mathrm{obs}$.

\subsubsection{The sizes of the clouds}

The sizes of the detected clouds, i.e. the number of on-cloud pixels, were compared to the sizes of the corresponding input clouds. The observed sizes are smaller than input sizes with an average factor of 0.85. The ratio remains remarkably constant, as 90 \% of all clouds in all the models have ratios between  $0.7 \dots 1.0$. Thus, if the cloud is detected using near-infrared excesses, it is likely to be detected close to its full extent, and only the most diffuse outer edges remain undetected.

\subsubsection{The apparent extinction}  

\begin{figure}
   \centering

   \includegraphics[width=0.99\columnwidth]{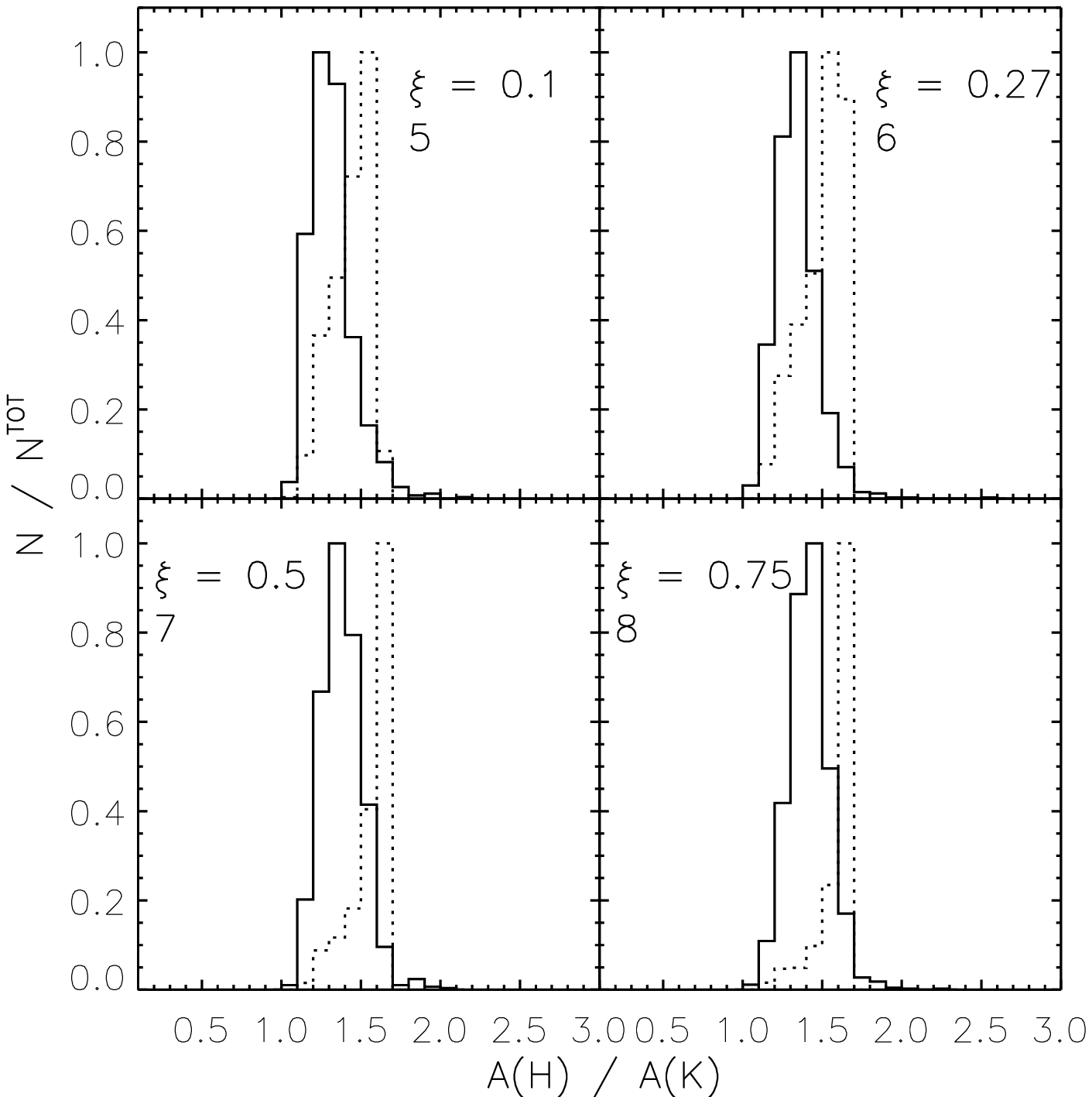} 
   \includegraphics[width=0.99\columnwidth]{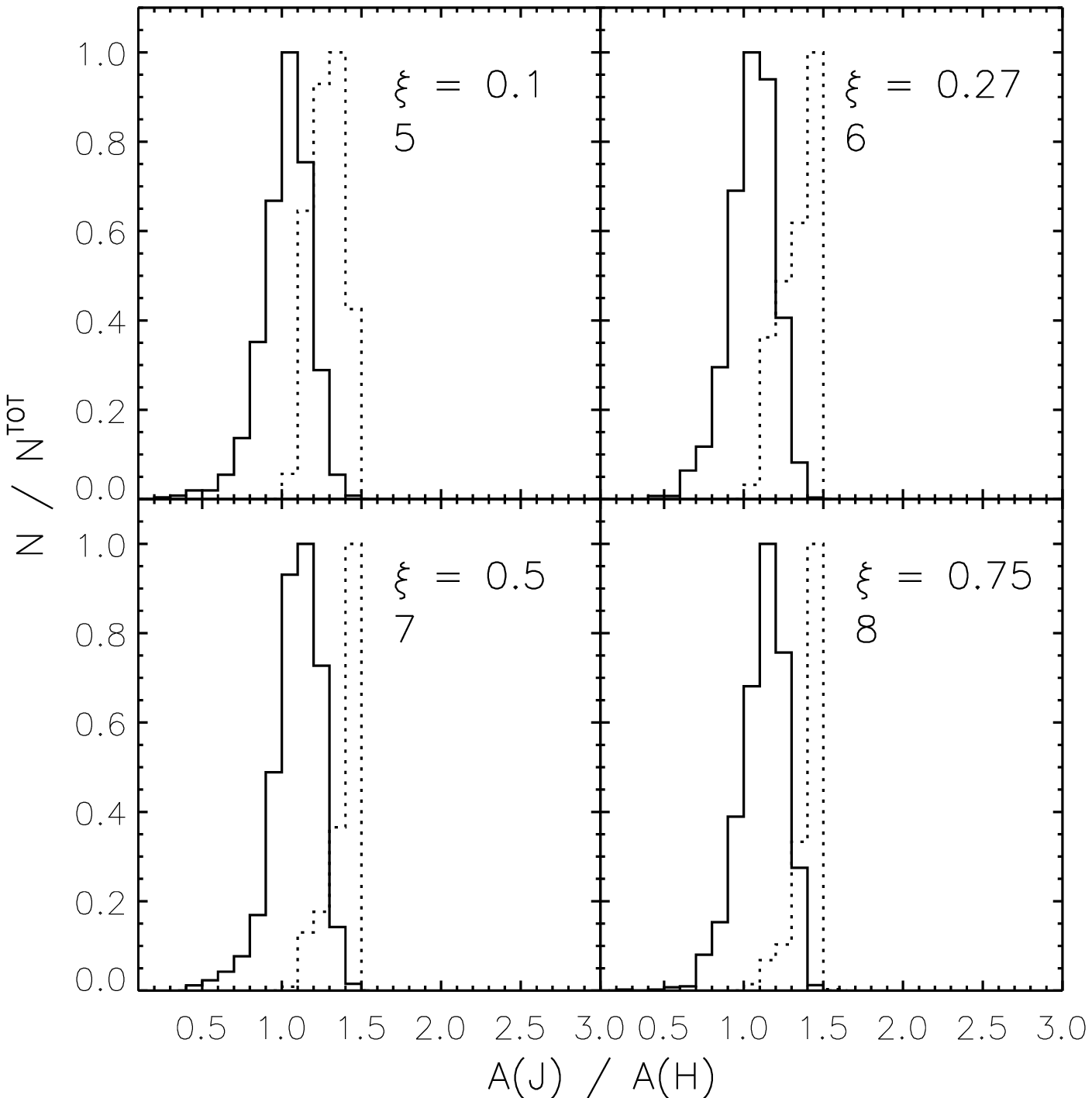} 

      \caption{ {\bf Top: }The histogram of the ratio of apparent extinction, $A^\mathrm{a}_\mathrm{H} /A^\mathrm{a}_\mathrm{K}$, evaluated in each on-cloud region. The apparent extinction is as defined by Eq. \ref{eq_apparent_A}. The data are for models 5-8, with the model number and the corresponding $\xi$ values are marked into the frames. The dotted lines show the corresponding histograms if the scattered light is neglected. (see section \ref{sec_results-first}) {\bf Bottom: }The same for $A^\mathrm{a}_\mathrm{J} / A^\mathrm{a}_\mathrm{H}$.}

         \label{fig_apparent_extinction}
   \end{figure}

As the absolute value of background flux in our models is known (the background flux is constant and the positions with zero optical depth are known), we can examine the \emph{apparent extinction} in the models. The apparent extinction is defined as:
\begin{equation}
A^\mathrm{a}_{(\lambda)} = -2.5 \log \frac{ F_{(\lambda)} }{F^{ \mathrm{bg} }_{(\lambda)} }
\label{eq_apparent_A}
\end{equation}
where $F^\mathrm{bg}_{(\lambda)}$ is the unattenuated background flux. The values of apparent extinction are typically in the order of a few tenths of a magnitude in our models. We estimate the wavelength dependency of the apparent extinction by calculating the ratios $A^\mathrm{a}_\mathrm{J}/A^\mathrm{a}_\mathrm{H}$ and $A^\mathrm{a}_\mathrm{H}/A^\mathrm{a}_\mathrm{K}$ at the positions of each extracted cloud. These ratios describe how the extinction depends on wavelength in the particular geometry, and thus the curve defined by them is often called an \emph{effective extinction law}. We calculate the ratios in each cloud by taking the average of ten pixels where the $E_\mathrm{H-K}$ is highest. The resulting histograms are shown in Fig. \ref{fig_apparent_extinction}. The distributions of $A^\mathrm{a}_\mathrm{H}/A^\mathrm{a}_\mathrm{K}$ values have their maxima between $\sim 1.35 \dots 1.55$, depending slightly on the value of $\xi$. The distributions of $A^\mathrm{a}_\mathrm{J}/A^\mathrm{a}_\mathrm{H}$ have their maxima only slightly above unity, typically at $\sim 1.1$. The ratios are clearly lower than in the case of foreground screen extinction (in our models: $\tau_\mathrm{J}/\tau_\mathrm{H}=1.58$ and  $\tau_\mathrm{H}/\tau_\mathrm{K}=1.65$, Draine \cite{draine03}).

\subsection{The amount of undetected mass}     
\label{sec_results-last}

\begin{figure}
\centering
  \includegraphics[width=1.0\columnwidth]{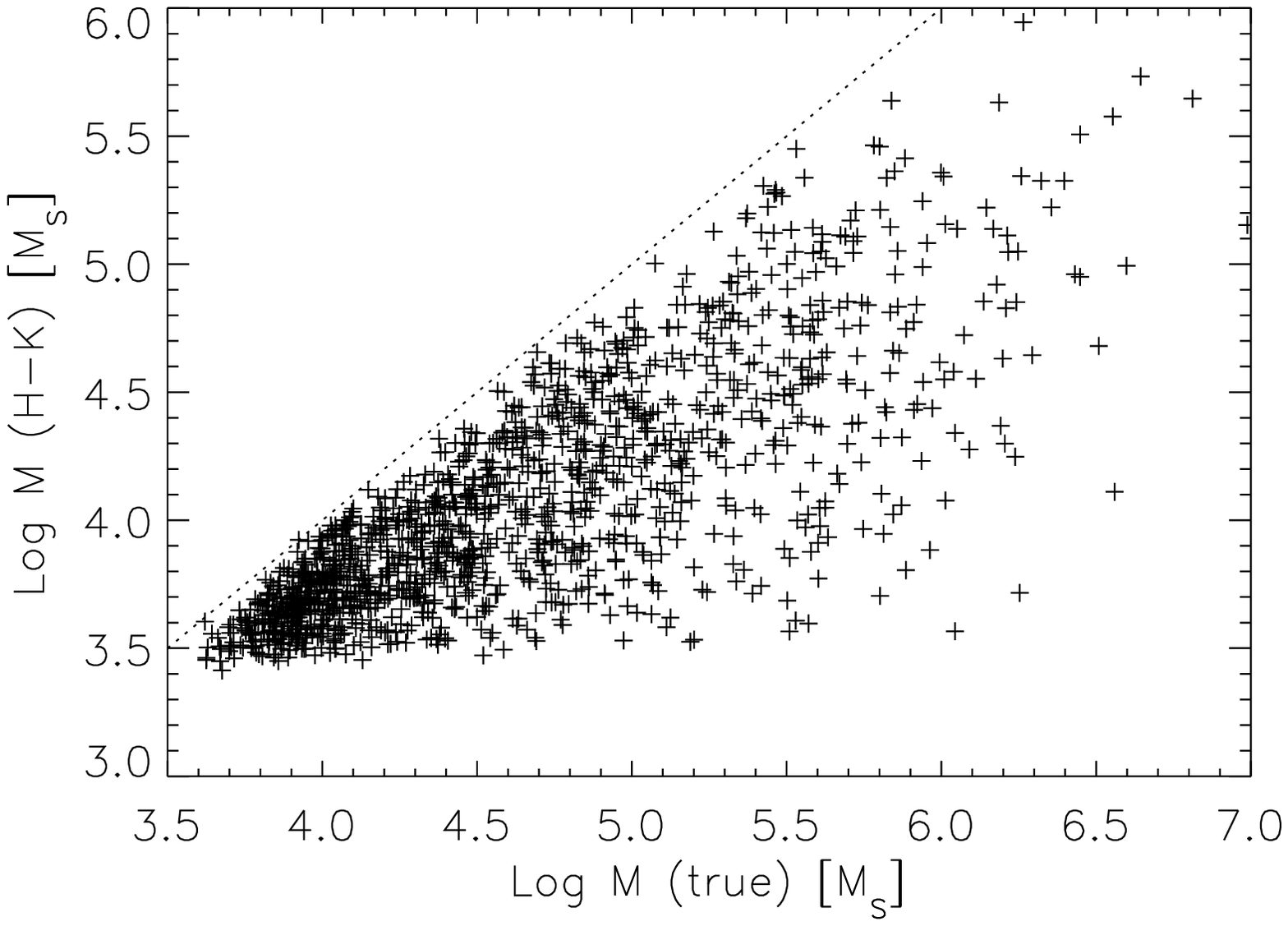}
  \includegraphics[width=1.0\columnwidth]{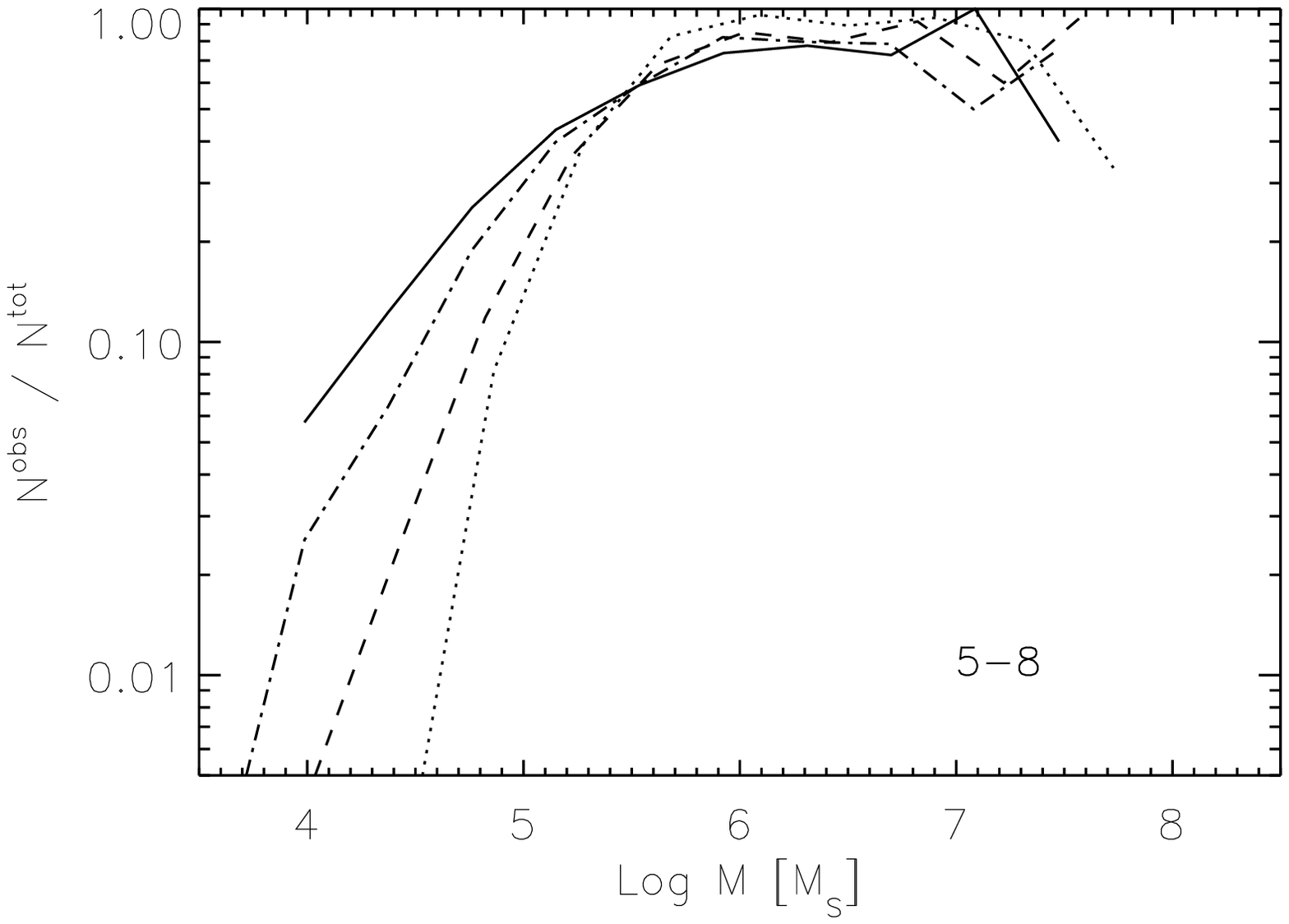}
  \caption{{\bf Top: }The observed masses of clouds as a function of the true mass of clouds in model 8. The observed masses are calculated from the $E_\mathrm{H-K}$ values by assuming the foreground screen geometry, and Galactic gas-to-dust ratio. The dotted line shows the one-to-one ratio. {\bf Bottom: }The fraction of detected clouds, i.e. the completeness function, in models 5-8 (dotted, dashed, dash-dotted, and solid lines). The masses are given in Solar masses.}
  \label{fig_obs-vs-true-clouds}
\end{figure}

As was already found in Paper I, the masses of the clouds that are detected can be drastically smaller than their true masses. As an example, Fig. \ref{fig_obs-vs-true-clouds} shows the comparison of the true and observed masses of clouds in model 8. The scatter in the relation is very high. Essentially, a cloud can be observed with any mass smaller than its true mass with a significant probability, down to the detection limit. In addition to the mass that is missed due to the underestimation of the masses of detected clouds, a fraction of clouds always remains completely undetected. This is mainly due to the population that is located on the further side of the galaxy. As the flux of the on-cloud pixels is completely dominated by the foreground flux, these clouds do not produce significant reddening features to the color maps. We calculate the total amount of undetected mass by summing up the observed extinction maps, and comparing them to the total sum of input column density maps. The resulting ratios, reported in column 10 of table \ref{tab_results}, are obviously very dependent of what is the dynamical range of input masses. Thus, the ratios given should only be taken as estimates of the order of undetected mass. 


Following the treatment in Paper I, we also calculated the fraction of detected clouds as a function of true mass by comparing the positions of clouds detected from input column density maps to the positions of clouds detected from the NICER extinction maps. The behaviour of the resulting completeness function was described in paper I. The larger set of models in this paper allows the examination of $\xi$ dependence of the completeness function. Fig. \ref{fig_obs-vs-true-clouds} shows an example of the $\xi$ dependence in models 5-8. The decrease in the completeness below $10^{5.5} $M$_\odot$ is stronger for smaller values of $\xi$. In the models with $\xi=0.1$ basically no clouds with $M<10^{4.5}$ M$_\odot$ are detected. In the models with $\xi=0.75$ clouds with $M=10^4$ M$_\odot$ still have a probability of a few percent to be detected. Above $10^{5.5}$ M$_\odot$ the detection probability is slightly higher for the models with lower $\xi$. This is simply due to the fact that the amount of clouds for which the fraction of foreground flux is high is larger in models with high $\xi$. The color of foreground light dominates the colors observed on these clouds over the small fraction of reddened background light.

\subsection{The significance of the scattered light}
\label{sec_results-first}

 During the simulations the scattered and emitted fluxes were registered separately, so we can explore the role of the scattering in the effective reddening and attenuation laws by using only the emitted fluxes. Figs. \ref{fig_reddening-law} and \ref{fig_apparent_extinction} show the corresponding reddening and attenuation laws with dashed lines. As the scattered light is not present the dispersion in histograms is much smaller. The peaks are clearly shifted towards higher ratios in both reddening and attenuation data. Thus, the scattered flux has a significant effect on both values.

\begin{figure}
  \centering

  \includegraphics[width=.95\columnwidth]{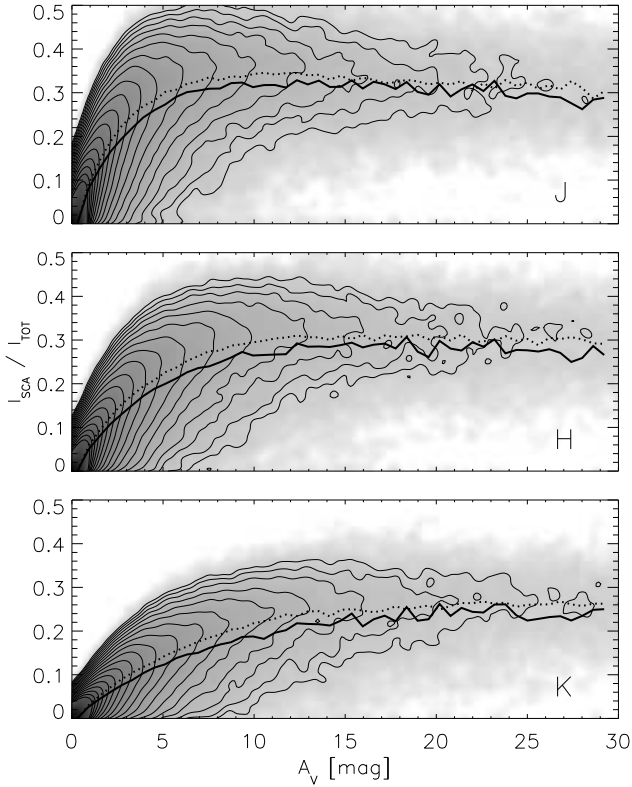}

  \caption{The density contours of the fraction of scattered light in model 5 at {\bf Top: }$J$, {\bf Middle: }$H$, and {\bf Bottom: }$K$ bands. The fraction refers to the scattered flux in the pixel divided by the total flux in that pixel (i.e. not by the total unattenuated flux). The solid lines show the median value of the relation for different $A_\mathrm{V}$ values in model 5 ($\xi=0.1$). For comparison, the dashed line shows the median ratio for model 8 ($\xi=0.75$).}

  \label{fig_isca-vs-trueav}
\end{figure}

The level of the scattered flux with respect to the total observed flux is shown in Fig. \ref{fig_isca-vs-trueav}. To better visualize the data, the figure shows the probability density of data points smoothed with a gaussian kernel. The data is shown for the model 5, and the overplotted lines show the median value of the relation in models 5 and 8. 
The increase of the fraction of the scattered flux is close to linear up to $A_\mathrm{V}\sim 3\dots5^m$. At higher optical depths the fraction saturates quickly, reaching its maximum at $A_\mathrm{V}\sim 10^m$ in $J$ and $H$ bands. In $K$ band the fraction continues to increase slowly until $A_\mathrm{V}\sim 30^m$. The value of $\xi$ has a relatively small, yet detectable, effect to the median value of the relation. The median value decreases slightly with increasing $\xi$, which results from the average radiation field experienced by a cloud being stronger when $\xi$ value is low.

In some studies an estimation of the optical depths along the line of sight has been attempted by using measurements of apparent extinction at two or more wavelengths (e.g. Howk \& Savage \cite{howk97}, \cite{howk00}). In the approach the scattered flux is neglected, and the resulting equations for radiative transfer become possible to solve. We applied this method to our models, but the equations failed to converge for most of the clouds. This failure was clearly due to the scattered flux, as the same procedure using only emitted fluxes indeed yielded improved estimates of optical depths. We also made an effort to use the data presented in Fig. \ref{fig_isca-vs-trueav} to solve numerically the actual optical depth. Typically, this operation multiplied cloud masses with a factor varying from $1 \dots 5$. Even though on average the cloud masses were thus improved, the effect was highly variable, and in every model the form of the mass function was completely destroyed. Considering that in our models the uncertainty caused by unknown background flux is neglible, we conclude that this correction is not feasible to gain an improved estimate of the mass function.

\section{Discussion}    %
\label{sec_discussion}


As has been well established in the previous studies of the subject, the large scatter and strong saturation of colors as a function of the total optical depth are dominating the simulations. These effects, originating from the varying relative geometry and inhomogeinity of dust distribution, affect both the level of extinction, and its wavelength dependency (i.e. the effective extinction law). The knowledge of realistic effective extinction law is crucial for numerous applications, where the observed signal from an object or region of interest needs to be corrected for the effects of dust. For example, for this purpose Calzetti et al. (\cite{calzetti97}, \cite{calzetti00}, \cite{calzetti01}) have derived an empirical attenuation laws for central regions of starburst galaxies. Likewise, theoretical attenuation curves have been calculated for several geometries of inhomogenous dust and stars (e.g. Pierini et al. \cite{pierini04} and references therein). In this paper we particularly examine the on-cloud pixels rather than the integrated light over a region covered with a mixture of GMCs and low column density lines of sight. We emphasize how the reddening properties of discrete cloud regions can vary, even in the fixed face-on geometry, when embedded GMCs are considered.



As we are considering only the on-cloud pixels, the effects due to the scattered flux undoubtly increase. In the most massive clouds, which are most likely to be detected in extinction maps, the fraction of the scattered flux to the total flux of a pixel can go up to several tens of percents (Fig. \ref{fig_isca-vs-trueav}). The scattered flux has a significant effect to the scale of color-excesses, dampening the overall color-excesses level with a factor of $~1.2 \dots 1.7$. 

Our simulations predict the effective reddening law, constructed from the values of the on-cloud pixels of detected clouds, to be close to unity. This result is valid only for the selected geometry, i.e. for a face-on galaxy. Even though the same average relation was found in all the models, we showed that when each cloud is examined separately, a $\xi$ dependence can be found (Fig. \ref{fig_E-histograms}). This demonstrates how a relatively straightforwardly measured observable reflects the geometry of the observed region despite the complicating effect of the scattered flux. In practice the determination of $E_\mathrm{J-H}/E_\mathrm{H-K}$ ratio for a single cloud is rather uncertain due to the photometric accuracy and generally small values of $E_\mathrm{J-H}$ and $E_\mathrm{H-K}$, and thus a sample of hundreds of clouds is likely to be needed to gain an estimate of the $\xi$ value of a face-on galaxy. It is interesting that even in the models with lowest average foreground emission, i.e. models with $\xi=0.75$, the effective reddening law remained very flat. Thus, if the overall conditions do not differ significantly from the simulations, high $E_\mathrm{J-H}/E_\mathrm{H-K}$ ratios should not be observed unless the foreground emission in the observed region really is neglible. This means that if a cloud region is observed to have a high $E_\mathrm{J-H}/E_\mathrm{H-K}$ ratio, it can be argued that the foreground screen approximation is valid for it.


In the case of an embedded cloud the effective reddening law does not relate the observed color-excesses directly to apparent extinction (as it would in the case of a foreground cloud). Thus, the effective extinction law cannot be inferred from those data alone. We evaluated the effective extinction law by determining the apparent extinction of on-cloud pixels, which is trivial to do in our models. We do note that this is not typically the case with spiral galaxies, as the knowledge of absolute background flux to a good precision would require a detailed model of the distribution of stellar light. The typical ratios of $A^\mathrm{a}_\mathrm{J}/A^\mathrm{a}_\mathrm{H}=1.15$ and  $A^\mathrm{a}_\mathrm{H}/A^\mathrm{a}_\mathrm{K}=1.35 \dots 1.55$ were found (Fig. \ref{fig_E-histograms}). The ratios are similar to the values derived in a pair of occulting galaxies by Berlind et al. \cite{berlind97}. We find this similarity quite convincing, as the regions sampled in Berlind et al. are relatively small, not much larger than the sizes of individual GMCs. Thus, the study might represent the conditions described by our simulations relatively well.


In addition to the description of reddening effects, we examine the mass functions derived from the color-excess data, and thus expand the work done in Paper I by increasing the number of models. The main emphasis regarding the mass functions is to examine the observed slope at different values of $\xi$. In the context of cloud mass functions, the direct consequence of the saturation of color-excesses is that the observed mass function is confined to a narrow dynamical range compared to the original mass function. A comparison of models which are otherwise similar, but have different $\xi$ values, shows that the confinement depends slightly on $\xi$. The difference in the high-mass end between the models with $\xi=0.1$ and $\xi=0.75$ is typically $\log M \sim 0.4$ M$_\odot$, while the low-mass end is at $\log M \sim 3.5$ M$_\odot$ in all the models.

In most of our models the observed slope of the mass function differs at least 0.3 from the slope of the true mass function. As is shown if Fig. \ref{fig_a_in_vs_a_out}, there is practically no correlation between the true and observed slopes. The observed slope depends obviously on the parameter $\xi$, as in the models with $\xi=0.1$ the slopes are generally $\gtrsim 2.5$, and in the models with $\xi=0.75$ they are closer to 2. However, $\xi$ clearly is not the only parameter affecting the slope. Whereas the models with $M_\mathrm{low}=10^4$ M$_\odot$ show practically constant observed slopes for all true slopes (blue plus signs in Fig.\ref{fig_a_in_vs_a_out}) , the slopes in the models with $M_\mathrm{low}=10^5$ M$_\odot$ do increase with the input slope in the cases where $\xi$ is large (green asterisks). The reason for this is most likely that the dispersion between the true and observed masses of low mass clouds is higher than in high mass clouds. Further, when the $\xi$ value is high the reddening signatures are stronger and the cloud definition better. When both $\xi$ and $M_\mathrm{low}$ are high, the relation between the true and observed masses is significantly better than for low $\xi$ and $M_\mathrm{low}$. Thus, if there is no reason to assume a truncation on the low-mass side of the mass function, the presence of lowest mass clouds makes the slope of the observed mass function quite insensitive to the true slope. However, if the truncation occurs at significantly lower masses than $10^5$, only a relatively small number of \emph{detected} clouds will be affected by it, and thus the cut-off should be meaningless regarding the observed mass function.


The near-infrared color-excess values, used in conjunction with the simple foreground screen approximation, do not provide a feasible method to determine the mass function of extra-galactic GMCs. However, the information carried by color-excess data regarding the geometric parameters remains interesting as such, and its sensitivity to dust features makes it a powerful tracer of GMCs despite the poor correspondance to real mass. Also, the results of this study are restricted to a face-on geometry where the most important factor, the fraction of foreground flux, is high and more importantly \emph{unknown}. Several situations can be imagined, where the distribution of GMCs with respect to the main illumination source is much better constrained (e.g. nuclear-centered rings, dust structures in highly inclined systems, etc.). The use of NIR color-excess as a mass tracer in these cases may prove to be feasible, but that remains to be confirmed with simulations specific for the geometry in question.

\section{Conclusions}   %
\label{sec_conclusions}

We have examined the near-infrared reddening properties of extra-galactic GMCs with Monte Carlo radiative transfer models. Our models resemble the face-on geometry and correspond to the dimensions of the most nearby spiral galaxies. We have investigated the effect of the relative scale height of GMCs, $\xi$, to the observed color-excesses and the mass spectrum of GMCs. We also derive the values of effective reddening law, $E_\mathrm{J-H}/E_\mathrm{H-K}$, and the effective extinction law for our models. The main conclusions of our simulations are as follows:

   \begin{enumerate}


      \item The effective NIR reddening-law of GMCs in a face-on geometry is predicted to have an \emph{average} slope close to unity. The observed $E_\mathrm{J-H}/E_\mathrm{H-K}$ ratios of on-cloud regions depend heavily on the exact geometry of the system, which in the context of a face-on galaxy is reflected by the value of $\xi$. From the simulations we find the average ratios of $E_\mathrm{J-H}/E_\mathrm{H-K}$=0.6, 0.7, 0.9, and 1.1 for the $\xi$ values of 0.1, 0.27, 0.5, and 0.75, respectively. The ratios towards spatially separate GMCs should be distinguishable within the typical photometric errors, and thus the value can be used as an indicator of line-of-sight geometry.

      \item The apparent extinction of GMC regions shows very flat wavelength dependency in $J$, $H$, and $K$ bands. We find that typically $A_\mathrm{J}^\mathrm{a}/A_\mathrm{H}^\mathrm{a}\sim 1.15$ and $A_\mathrm{H}^\mathrm{a}/A_\mathrm{K}^\mathrm{a}\sim 1.35 \dots 1.55$, the latter depending slightly on $\xi$. 

      \item The comparisons of the models with and without scattering suggests that the scattered flux has a non-neglible effect to the color-excesses and the apparent extinction of GMC regions.


      \item We extend the results of Paper I, confirming that near-infrared color-excess is an unrelieable measure of the column density of GMCs. If $E_\mathrm{H-K}$ values are transformed to masses assuming the Galactic reddening law, \emph{on average} a fraction in the order of 10-20\% of the total mass is recovered. In the case of individual GMCs the percentage can vary alot, being basically anything between $0\dots \sim 50$\%. 

      \item The mass function constructed by using the color-excesses and the Galactic reddening law is sensitive to the parameter $\xi$. The models with smaller $\xi$ values tend to have steeper ($\gtrsim 2.5$) slopes than models with high $\xi$ values ($\gtrsim 2.0$). The dynamical range of the underlying mass function also modifies the observed slope. There is a strong degeneracy in the observed slope with respect to the parameters $\alpha_{true}$, $\xi$, and $M_\mathrm{low}$. As none of these parameters is observationally well constrained, it is not possible to determine $\alpha_{true}$ based only on a measurement of $\alpha_{obs}$.

   \end{enumerate}

\begin{acknowledgements}

We thank the referee, S. Bianchi, whos comments improved the paper greatly.

\end{acknowledgements}

\end{document}